\documentclass[lettersize,journal]{IEEEtran}
\usepackage{amsmath,amsfonts}
\usepackage{algorithmic}
\usepackage{array}
\usepackage[caption=false,font=normalsize,labelfont=sf,textfont=sf]{subfig}
\usepackage{textcomp}
\usepackage{stfloats}
\usepackage{url}
\usepackage{verbatim}
\usepackage{graphicx}
\def\BibTeX{{\rm B\kern-.05em{\sc i\kern-.025em b}\kern-.08em
    T\kern-.1667em\lower.7ex\hbox{E}\kern-.125emX}}
\usepackage{balance}
\usepackage[hidelinks]{hyperref}
\newcommand{\orcid}[1]{\href{https://orcid.org/#1}{\includegraphics[width=8pt]{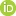}}}

\newcommand{\etal}{\textit{et al.} }

\begin{document}


\title{On-Demand Myoelectric Control Using Wake Gestures to Eliminate False Activations During Activities of Daily Living}

\author{Ethan~Eddy\orcid{0000-0002-8392-3729},~\IEEEmembership{Member,~IEEE,} Evan~Campbell\orcid{0000-0001-5399-4318},~\IEEEmembership{Member,~IEEE,} Scott~Bateman\orcid{0000-0003-3592-2163}, \\ and~Erik~Scheme\orcid{0000-0002-4421-1016},~\IEEEmembership{Senior~Member,~IEEE}
\thanks{{E. Eddy, E. Campbell, and E. Scheme are with the Institute of Biomedical Engineering and the Department of Electrical Engineering, University of New Brunswick, Fredericton, Canada;} {\tt\footnotesize eeddy@unb.ca, evan.campbell1@unb.ca, escheme@unb.ca}.}
\thanks{{S. Bateman is with the HCI and SPECTRAL Labs, University of New Brunswick, Fredericton, Canada;} {\tt\footnotesize scottb@unb.ca}
}
\thanks{*This work has been submitted for possible publication. Copyright may be transferred without notice, after which this version may no longer be accessible.}
}


\maketitle

\begin{abstract}
While myoelectric control has recently become a focus of increased research as a possible flexible hands-free input modality, current control approaches are prone to inadvertent false activations in real-world conditions.
In this work, a novel myoelectric control paradigm---on-demand myoelectric control---is proposed, designed, and evaluated, to reduce the number of unrelated muscle movements that are incorrectly interpreted as input gestures .
By leveraging the concept of wake gestures, users were able to switch between a dedicated control mode and a sleep mode, effectively eliminating inadvertent activations during activities of daily living (ADLs).
The feasibility of wake gestures was demonstrated in this work through two online ubiquitous EMG control tasks with varying difficulty levels; dismissing an alarm and controlling a robot.
The proposed control scheme was able to appropriately ignore almost all non-targeted muscular inputs during ADLs ($>$99.9\%) while maintaining sufficient sensitivity for reliable mode switching during intentional wake gesture elicitation.
These results highlight the potential of wake gestures as a critical step towards enabling ubiquitous myoelectric control-based on-demand input for a wide range of applications.
\end{abstract}


\section{Introduction}  
With new opportunities emerging with mixed reality, smart homes, and other connected technologies, modern-day human-computer interaction is rapidly evolving \cite{21_cent_comp}. 
As we increasingly engage within ubiquitous computing environments (i.e., computing anywhere and everywhere), there is a renewed need for always-available inputs that are both reliable and robust.
Myoelectric control, which involves the recognition of muscle-based inputs (or gestures) via the electromyogram (EMG) signals generated during muscular contractions \cite{emg_hms}, is an attractive solution due to its potential intuitiveness, subtlety, and ability to work in real-world scenarios where other technologies may not \cite{eddy_hci}.  
By leveraging machine learning to extract distinctive patterns of muscle activity to predict gestural intent \cite{scheme_electromyogram_2011}, EMG has the potential to unlock many interesting and unique interaction opportunities. 

Despite its history of success within the prosthetics community \cite{oldest_prosthetic1}, the use of myoelectric control for human-computer interaction (HCI) as a general-purpose input modality only materialized in the early 2000s \cite{saponas_demonstrating_2008}.
In this early exploration period, most HCI researchers embraced the best practices for myoelectric control as established over decades by the prosthetics community \cite{eddy_hci}. 
The issue, however, is that these control schemes, initially designed for the continuous micro-adjustment of a prosthetic limb, fundamentally differ from those required for general-purpose ubiquitous use cases. 
Correspondingly, there has been a prevailing perception that myoelectric control lacks the robustness required for real-world general-purpose use, often being characterized as an ``imprecise'' input modality \cite{imprecise_but_fun}.

A fundamental distinction between myoelectric control for powered prostheses and general-purpose input is the difference between \textit{dedicated} and \textit{on-demand} input.
Dedicated control refers to a system that is always classifying EMG as input intent, often with an `other' or `no movement' class acting as a `do nothing' command. Controlling a prosthetic device often assumes dedicated  \textit{continuous control}, where individuals constantly adjust their prosthesis based on predictions occurring on milliseconds worth of data \cite{continuous_control}. In these cases, the classification problem is considered to be \textit{closed-set}, where the EMG inputs to the system are assumed to correspond to one of the $N$ classes used for training.
The closed-set continuous control assumption is a seemingly appropriate approach for prosthesis control because amputees primarily engage in dedicated single-task device control; that is, it is assumed that they only activate the muscles of their residual limb to control the prosthesis.
In turn, the prosthesis is always on, ready to be controlled, and there is no other task to be performed. 
This is not always the case for a user of a `general purpose' myoelectric control system, where intermittent control is often desirable. For example, imagine controlling a slideshow with an EMG device and stopping to type on the keyboard to bring up a website to show the audience. In this situation, typing on the keyboard might produce muscle inputs that are unexpected in the closed-set, inadvertently advancing the slide deck. So, a dedicated control scheme in these settings precludes the ability for users to perform additional activities without worrying about inadvertent activations, an essential requirement for any ubiquitous input technology.

A highly desirable characteristic of general-purpose myoelectric control would be to enable situational input in ubiquitous environments. For example, controlling a music player while walking to work, subtly accessing information on a mixed-reality display while in the grocery store, or dismissing a phone call while driving \cite{libemg}.
However, the dedicated and continuous myoelectric control schemes previously explored in the HCI literature (e.g., \cite{saponas_demonstrating_2008,drone,smart_garments}) have struggled to offer such capabilities.
To realize the required level of robustness, on-demand myoelectric control may offer promise through the intentional activation and deactivation of a dedicated control mode. 
Toward this goal, an on-demand myoelectric control system based on \textit{wake gestures} is proposed, designed, and evaluated.
Similar to the now familiar wake words ``Hey Google'' or ``Hey Siri'', a widely familiar gesture---a finger snap---is identified as a toggle to easily transition between modes.

Leveraging the added temporal information encoded in dynamic sequences of muscle contractions, these patterns are better differentiated from other patterns of EMG elicited during ADLs \cite{eddy_hci}. In this work, a straightforward Dynamic Time Warping (DTW) \cite{dtw} based template matching approach was found to enable the rejection of inadvertent muscle inputs while remaining largely resilient to possible variations in gesture dynamics. 

To evaluate the proposed on-demand myoelectric control approach, 20 individuals with varying degrees of experience with EMG used the system during various activities of daily living (ADLs), including walking, writing, driving, typing, and phone use. While performing the ADLs, participants were randomly cued to complete a secondary task using EMG as control (either using a gesture to dismiss an alarm or controlling a remote-controlled robot around an obstacle course). 
The proposed system ignored $>$99.9\% of all inadvertent inputs during these tasks while reliably recognizing the wake gesture when intentionally activated.
These results suggest a promising path forward for myoelectric control as a robust, on-demand, general-purpose ubiquitous input modality.

\section{Related Work}

\subsection{Myoelectric Control}
Since the 1960s, myoelectric control has primarily evolved to enable amputees to control powered prostheses \cite{oldest_prosthetic1,eddy_hci}. The early conventional control schemes took a one-muscle, one-function approach, where individual antagonistic muscles controlled a single degree of freedom (e.g., opening and closing the prosthesis) \cite{direct_control}. While simple and practical, the input space of these systems was limited, requiring co-contractions (i.e., the co-activation of both antagonistic muscles) to switch the prosthesis mode (e.g., switching from hand open/close mode to wrist flexion/extension mode). Not only was this cognitively and physically cumbersome, but it often resulted in inadvertent false activations---leading to user frustration and device abandonment \cite{abandonment}.  

In 2003, Englehart \etal set the standard for the more intuitive, dedicated continuous pattern-recognition-based myoelectric control systems \cite{continuous_control} that are adopted today worldwide, both commercially for prosthesis control and for the majority of EMG-based gesture recognition research \cite{eddy_hci}. 
In these control systems, the EMG signal is first filtered to remove unwanted noise, windowed (i.e., split into frames with a predefined length and increment that governs the controller's responsiveness), features are extracted, and finally, a classifier continuously outputs a class (i.e. gesture) prediction.
Although these systems enable the recognition of a diverse set of inputs, they are prone to false activations when out-of-set gestures are elicited \cite{out-of-set-emg}.
While some have sought to limit the impact of unwanted inputs through techniques like rejection \cite{rejection,rejection2}, current myoelectric control systems still assume a closed-set environment, forcing user behaviours into a dedicated input mode.
In turn, the adoption of these dedicated continuous control schemes for other applications in HCI has been misguided. 

\subsection{The False Activation Problem}
Myoelectric control began to receive interest within the general human-machine interaction community in the early 2000s following its history of success for prosthesis control \cite{eddy_hci}.
Since then, it has been explored for a variety of applications, including drone control \cite{drone}, pen augmentation \cite{pen_agumentation}, and human-robot interaction \cite{plug-and-play}, and many have worked to progress algorithms and approaches in offline settings \cite{campbell2020feasibility, campbell2021deep}.
Despite this extensive exploration, a generally overlooked critical issue is separating intentional (closed-set) inputs from unintentional (open-set) ones.
Generally, current systems suffer from the problem of all actions turning into device commands (akin to the "Midas Touch" problem \cite{midas_touch} whereby all actions, intentional or not, are associated with an input command).
In particular, this issue has been exacerbated by offline studies and simulated online evaluations that inadequately represent the real-world ubiquitous use of myoelectric control.
As highlighted by Chang \etal, this is particularly troublesome as there is a direct overlap in feature space between many commonly used control gestures (e.g., hand open/close and wrist flexion/extension) and typical activities of daily living (ADLs), meaning that false activations are inevitable with the current continuous control paradigm \cite{jason_adl}.

\subsection{Wake Words}
Similarly to myoelectric control, speech recognition struggles with unintended activations when the user is not deliberately engaging with the system.
This led to the adoption of wake words \cite{speech_wakewords}---a simple and unique input that can toggle or wake up the system when desired.
Despite debates on the optimality of wake words from the perspective of designing ideal interactive systems \cite{speech_wakewords}, they continue to be integrated into widely adopted commercial devices, such as the iPhone (Siri), Microsoft (Cortana), Google Assistant, and Amazon Alexa -- highlighting their practicality and real-world robustness.
Correspondingly, it is conceivable that leveraging such an approach could be advantageous for solving the false activation problem for myoelectric control \cite{rf-sensor-ww}.

While myoelectric control researchers have now acknowledged the challenge of out-of-set inputs \cite{out-of-set-emg}, very few have explored the concept of wake gestures.
Kumar \etal showed the potential of wake gestures for myoelectric control, but their pilot work was limited to offline data, and importantly, the out-of-set inputs used for validation were the other gestures collected in the study rather than ADLs, limiting its real-world generalizability \cite{ww_pradeep}.
Tavakoli \etal proposed the concept of pulse-based locking mechanisms, however, these were specifically designed for prosthesis control use-cases \cite{tavakoli_1,tavakoli_2}.
The commercial Myo Armband by Thalmic Labs leveraged a wake gesture (a double-finger tap) and an automatic two-second timeout to enable moded entry into the underlying gesture recognizer.
The Myo approach, however, arguably lacked the requisite robustness for real-world use \cite{myo_double_tap,locked_out,wake_gesture_problematic}, with many indicating that it was overly sensitive, lacked reliability, and was sometimes challenging to trigger\footnote{https://www.youtube.com/watch?v=Ns2VZaWA9CE}\footnote{https://www.pcmag.com/reviews/myo-gesture-control-armband}\footnote{https://www.tomsguide.com/us/myo-gesture-control-armband,review-2870.html}. 
Additionally, this device is now discontinued, and its design, implementation, and evaluation were never externally validated or publicly released, meaning there is currently no way to evaluate or baseline the wake gesture approach Myo took.
Correspondingly, this work seeks to build on and set the standard for what the Myo Armband started by designing, developing, and validating a framework that vastly improves the basis for using EMG as a ubiquitous input modality.

\begin{figure*}[h!]
    \centering
    \includegraphics[width=0.762\linewidth]{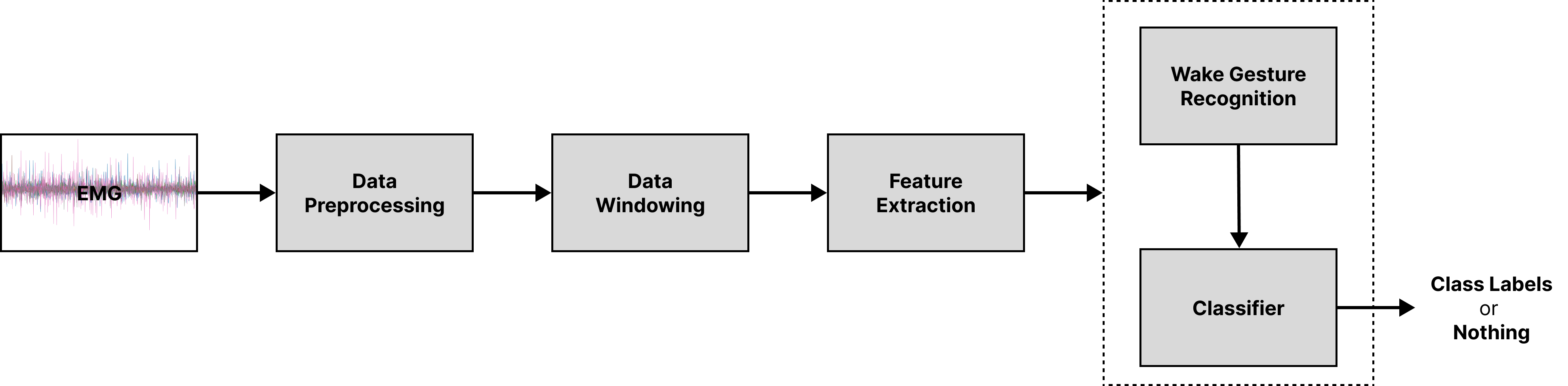}
    \caption{The myoelectric control pipeline for enabling on-demand input. The entire system is available in LibEMG, a toolkit for myoelectric control \cite{libemg}.}
    \label{fig:pipeline}
\end{figure*}

\section{Methods}

\subsection{The On-Demand Myoelectric Control Pipeline}
In the proposed on-demand myoelectric control system, the predictions from the underlying classifier (constantly being generated) are silenced until the wake gesture activates the system, changing it to an `input' mode.
Subsequently, an additional independent layer has been integrated into the pipeline (see Figure \ref{fig:pipeline}), facilitating the activation or deactivation of the output, thus affording the ability to ignore inadvertent device commands when desired.
Upon toggling the system into the input mode, the classifier predictions are sent to the device, enabling control until a subsequent deactivation event is recognized and the system is put back into its sleep state.
This deactivation could occur through a period of inactivity (the strategy of many smart home systems), through the detection of the wake gesture, or the detection of a separate sleep gesture. 

\subsection{Design Considerations for Wake Gesture Systems}
Compared to voice-activated wake words, EMG-based wake gestures present distinct challenges, as mastering precise underlying patterns of muscle contractions to improve the separability between gestures while maintaining sufficient repeatability is difficult.
For example, amputees often undergo lengthy training regimes to accurately control their prostheses, with some turning to serious games for learning \cite{aaron_momo}.
This challenge becomes even more apparent for general-purpose use due to differences between hand positions (or postures) and the elicited EMG. 
For example, a user could position the hand into an open posture but with very low levels of EMG, making it difficult for the underlying myoelectric control model to differentiate the gesture from the No Movement class. 
Some of the design considerations for EMG wake gestures are described below. 

\subsubsection{The Trade-off Between False Positives and False Negatives}
When designing wake gesture systems, there is a fundamental trade-off between false negatives and false positives.  
In general, a more restrictive system (i.e., requiring more confidence in a true positive) will make it potentially more difficult for the user to elicit a gesture that will be recognized, resulting in increased false negatives.
Although false negatives (missing intended inputs) are undesirable, wake gesture systems should be optimized to prioritize suppressing false positives. This is because, while false negatives require the user to re-elicit a command, false positives produce unintentional and potentially confusing actions or commands that impose additional time and attention costs to correct \cite{fn_v_fp}. For example, a robot could erroneously move in an unintended direction, or a phone call could be inadvertently ended---both of which require additional input to recover to the original state. 

\subsubsection{Intuitiveness and Social Acceptability}
In the work by Pomykalski \etal \cite{snap_motivation} that investigated the use of wake gestures via computer vision, the \textit{social acceptability} of the selected gesture was highlighted as a crucial consideration among participants.
Fortunately, compared to other input modalities like computer vision or speech recognition, myoelectric control's subtlety enhances its appropriateness since gestures can be made discretely in inconspicuous positions if desired, which is not often possible with modalities like accelerometers and computer vision.
Another important consideration is the \textit{effort and intuitiveness} of the chosen gesture.
Although longer gesture sequences or ``secret handshakes'' that chain multiple gestures could reduce both the potential false positives and false negatives, they come at the cost of prolonging system activation time and decreasing intuitiveness. 

\subsubsection{Repeatability and Separability}
The \textit{repeatability} of the gesture is a measurement of the similarity between two gesture templates, indicating a user's ability to consistently match the gesture.
While users are accustomed to controlling for this in speech recognition, it is arguably more difficult for muscle-based inputs which require synergistic muscle coordination and regulation through proprioception.
Similarly, the \textit{separability} of the gesture is a measure of its distance (or difference) from all other gestures and unrelated muscle activations.
Selected wake gestures must have appropriate separability in the underlying EMG feature space and adequate repeatability for consistent template matching. 
This should not be confused with kinematically separable gestures (or postures) such as those used in computer-vision-based systems, which may or may not elicit meaningful EMG activity.

\subsection{Participants and Hardware}
As approved by the Research and Ethics Board at the University of New Brunswick (on file as REB 2022-122), 20 participants (between the ages of 20 and 31) were recruited for this study. 
The participants had a range of experience with myoelectric control---with most having minimal to none.
The experiment lasted approximately 1.5 hours, and participation was entirely voluntary, with no rewards or incentives provided.

\begin{figure}[h!]
    \centering
    \includegraphics[width=0.325\linewidth]{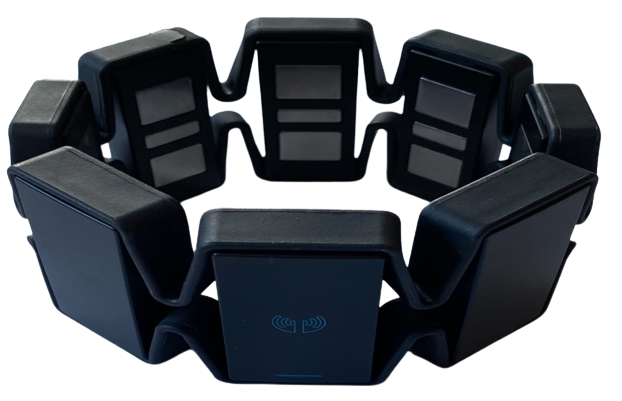}
    \caption{The Myo Armband, a previously commercially available surface-EMG wearable device.}
    \label{fig:myo}
\end{figure}

All data in this experiment were recorded using the Myo Armband (see Figure \ref{fig:myo}), a previously commercially available surface-EMG wearable device recording EMG at 200 Hz across eight channels.
The Myo Armband was chosen due to its ease-of-use, prevalence in the HCI literature, and its previous use within the consumer market.  
At the start of the session, participants placed the armband on their dominant forearm, at approximately 1/3 of the length of the forearm, proximal to the elbow. 
EMG data were recorded for the entire session to enable the post hoc analysis of the system's performance.

\subsection{Model Training}

\subsubsection{Continuous Myoelectric Control System}
\label{sec:classifier}
For each participant, a personalized, continuous myoelectric control model was trained to discriminate between five gestures: Wrist Flexion, Wrist Extension, Hand Open, Hand Close, and No Movement (Rest).
These gestures were chosen as they were also employed by the Myo Armband and are widely adopted in research \cite{smart_garments}.
This was the underlying classification model used for both tasks highlighted in Section \ref{sec:tasks}.
To train this model, participants were prompted to elicit the displayed gesture in a traditional screen-guided training session (i.e., where users follow along with images on a screen indicating which contraction they should be eliciting) \cite{prosthesis_guided_training}.
Five repetitions of moderate-intensity ramp contractions (i.e., contractions starting from rest) of three seconds each were captured for each of the five input classes. 


The underlying continuous classifier (i.e., the `classifier' block in Figure \ref{fig:pipeline}) was a Linear Discriminant Analysis (LDA) model---a simple machine learning model commonly used for prosthesis control \cite{campbell2019linear}.
All EMG data were split into windows of size 200 ms with 100 ms increments, resulting in ten classifier predictions per second.
Hudgins' Time Domain feature set (i.e., Mean Absolute Value, Zero Crossings, Waveform Length, and Slope Sign Changes) \cite{hudgins} was used to describe each window of EMG for classification.
A proportional velocity control scheme was also adopted, meaning that each class prediction was accompanied by an associated speed, as is common for prosthesis control \cite{motion_normalized}.

\begin{figure*}[h!]
    \centering
    \includegraphics[width=0.8\linewidth]{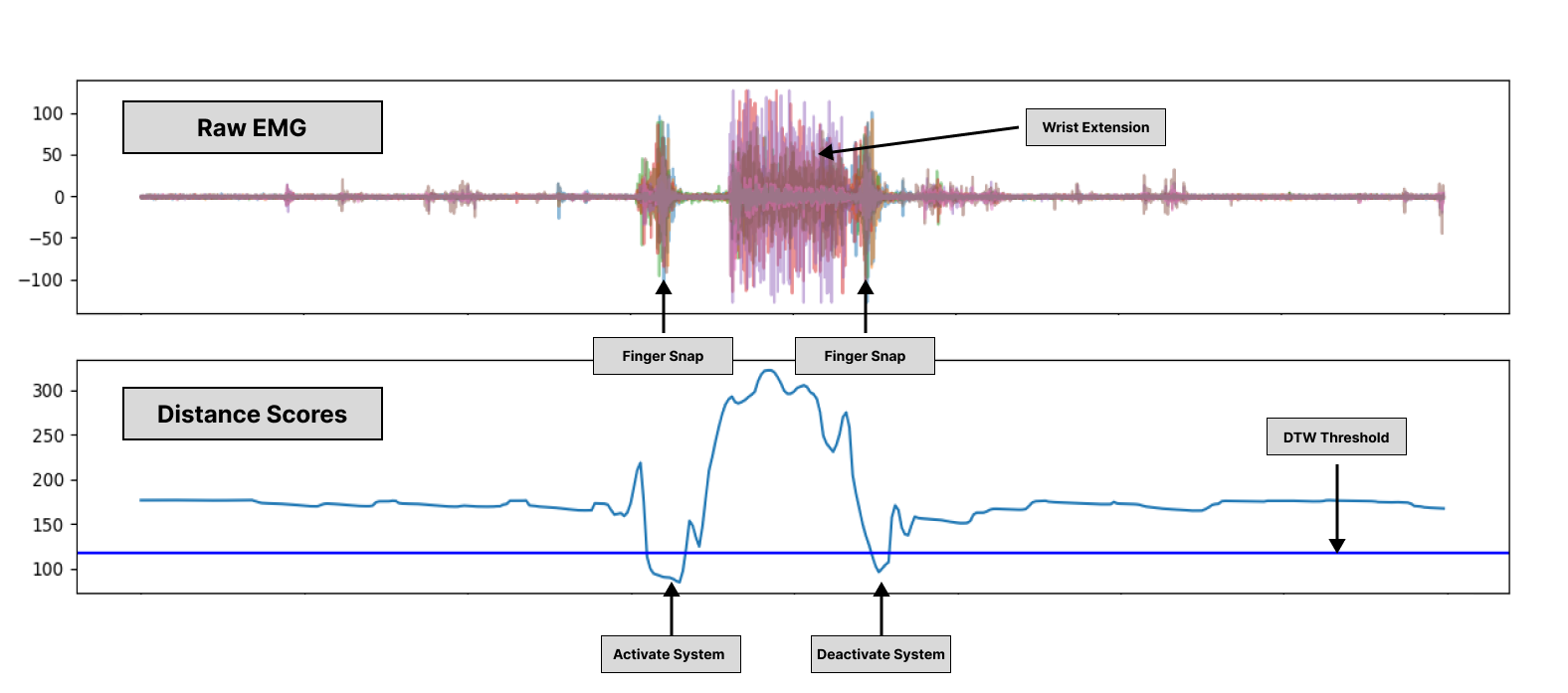}
    \caption{An example of a participant dismissing an alarm by unlocking the system with a finger snap, subsequently extending their wrist to dismiss the alarm, and snapping their finger again to turn the system off.}
    \label{fig:test}
\end{figure*}

\subsubsection{Wake Gesture Recognition System}
\label{sec:ww_train_sec}
The wake gesture recognition system (i.e., the `wake gesture recognition' block in Figure \ref{fig:pipeline}), which is independent of the aforementioned continuous control model, employed a simple Dynamic Time Warping (DTW) approach similar to that proposed in the offline study by Kumar \etal \cite{ww_pradeep}.
This approach was used as it enables the recognition of dynamic gestures by evaluating the Euclidean distance between two multi-variate time series (i.e., templates) after accounting for different warpings \cite{dtw}. It has been shown to have better wake gesture detection performance than non-temporally encoded algorithms because it enables small variations in the way that gestures are elicited \cite{ww_pradeep}.
As highlighted below, 20 ground-truth templates were recorded for each participant and used as target templates.
Each template (one second of raw EMG data) was split into windows of 150 ms with 50 ms increments as proposed in \cite{ww_pradeep} and confirmed via pilot studies. Compared to the continuous classifier, the shorter window length and increment were chosen to improve the temporal resolution of the DTW. Finally, the Root Mean Square (RMS) feature was extracted from each of the eight channels.

The threshold used for the DTW was computed as follows: 
\begin{equation}\label{eq:1}
    D = \sum_{i=1}^{N} dist(t_{1}, t_{2})
\end{equation}
\hspace{2mm}
where $N$ is the unique combination of all two templates, $t_{1}$ is one of the twenty templates, $t_{2}$ is another of the unique templates, $i$ is the current unique pair of templates, $dist$ is the DTW alignment cost between two multivariate time series, and $D$ is a vector of distances from all templates to the others.
\begin{equation}\label{eq:2}
    T_{thresh} = \frac{D}{N} + s \sigma
\end{equation}
\hspace{2mm}
where $T_{thresh}$ is the wake gesture model's threshold, $\sigma$ is the standard deviation of the vector $D$, and $s$ is the number of standard deviations away from the mean distance between templates allowed.\\

For the wake gesture recognition system, one-second templates were passed through the DTW model every 50 ms using a sliding window.
The average distance of each new template to all the gesture templates in the training set was computed and compared to the tuned threshold $T_{thresh}$.
If the average distance between the new template and the training set was below the predefined threshold, the template was classified as a finger snap gesture (toggling the control mode between sleep and input). 
A majority vote was then taken over five subsequent predictions (250 ms) to further improve the algorithm's robustness to false activations.
An example decision stream of the thresholds of incoming EMG templates is shown in Figure \ref{fig:test}.
Users were provided with an activation and deactivation audio stimulus (similar to the generated sound when plugging in and removing a USB stick from a computer) to indicate when they had successfully activated and deactivated the system.

To calibrate the wake gesture recognition system, participants were prompted to snap their fingers within a one-second data capture period for 20 repetitions.
The finger snap gesture was selected based on previous wake gesture work for computer vision \cite{snap_motivation}, initial pilot studies, and because finger snaps are inherently dynamic, playing into the strengths of the temporal nature of the DTW approach used for detection.

Following data acquisition, the wake gesture recognition model was tuned.
Participants were guided through a one-minute unstructured session where mock ADL data were captured to obtain the ideal distance threshold (see Equation \ref{eq:2}).
Additionally, participants were asked to elicit gestures that might incorrectly activate the system, such as tapping and waving their fingers.
This data was then used to generate the false positive rate (FPR) of a receiver operating characteristic (ROC) curve.
The point nearest the top left corner that had no false positives was chosen as the threshold, as this represented the model's best true positive rate (its ability to recognize the training templates as finger snaps) while ensuring a false positive rate of zero (its ability to ignore all data recorded during the one minute of mock data).

\subsection{Activities of Daily Living (ADLs)}
\label{sec:activities}
After training the system to recognize the wake gesture and the five continuous control gestures, it was evaluated during a representative set of ADLs.
The activities chosen to evaluate the control system came from conversations with pilot participants about the activities they spend the most time doing throughout the day. 
While these tasks do not represent all possible variations of everyday activities, they were meant to encompass various tasks whereby EMG-based inputs may be useful (e.g., a music player while walking or a gesture-based keyboard augmentation).
Two minutes of data were recorded for each activity, and each activity was repeated across both tasks for a total of 20 (2 minutes x 2 tasks x 5 activities) minutes of ADL data per participant.
These activities were presented randomly to the participants for each task.
The following is an overview of the five ADLs evaluated in this study:

\begin{itemize}
    \item \textbf{Walking:} Participants were asked to walk on a stationary treadmill at a slow walking speed of 2 km/hr. 
    \item \textbf{Writing:} Participants were given a book, paper, and a pen and were asked to write a passage while occasionally changing pages. 
    \item \textbf{Typing:} Participants were asked to type at a comfortable speed and perform an untimed typing test\footnote{https://official-typing-test.com/test/untimed.html}. 
    \item \textbf{Driving:} Participants used a Thrustmaster TMX \footnote{https://www.thrustmaster.com/en-us/products/tmx-force-feedback/} force feedback driving simulator (which vibrated during driving) and the RaceRoom Racing Experience game to simulate a realistic driving experience.
    \item \textbf{Phone:} Participants were asked to complete common phone activities including responding to text messages, scrolling through social media, and any other phone-related tasks.
    \item \textbf{Other:} Participants elicited a variety of other actions, such as standing up from their chairs, scratching their noses, grabbing their phones out of their backpacks, tying their shoes, talking with their hands, etc. Although these activities were not directly prompted, they ended up being additional tasks that further tested the resilience of the wake gesture algorithm. 
\end{itemize}

\subsection{Myoelectric Control Tasks}
\label{sec:tasks}
The proposed on-demand myoelectric control system was evaluated in two unique and fundamentally different tasks.
These tasks were evaluated in the context of each ADL presented above. 
The first task involved a complete context switch, as participants navigated a robot around an obstacle course. The second task represented the ubiquitous use of myoelectric control and involved dismissing an alarm. Users were periodically prompted to address each task, with notifications recurring four times within each two-minute activity of daily living at 30-second intervals. When prompted, users would stop the ADL activity they were doing to engage with the particular task. It is important to note that we were not concerned about participants' performance within the dedicated control tasks nor with the control enabled by the classifier, as this study was focused on establishing the on-demand control framework with a focus on the DTW-based wake gesture recognition system.
Correspondingly, the tasks were meant to serve as a realistic environment to evaluate the intentional activation of the wake gesture.

\begin{figure}[h!]
    \centering
    \includegraphics[width=0.7\linewidth]{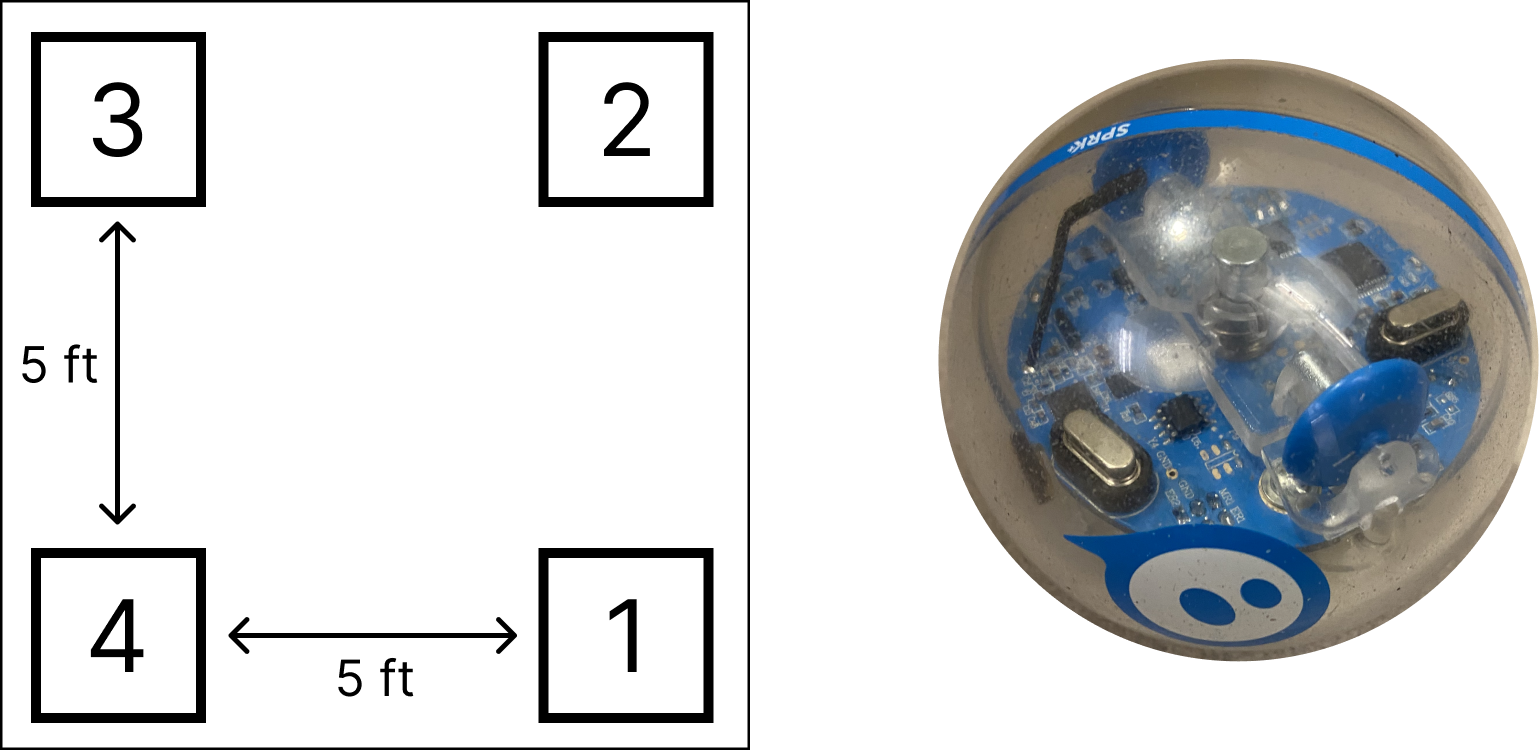}
    \caption{A representation of the Sphero task that participants were required to complete. The task was considered done when the participant moved the Sphero from targets 1 $\rightarrow$ 2 $\rightarrow$ 3 $\rightarrow$ 4 $\rightarrow$ 1.}
    \label{fig:sphero}
\end{figure}

\subsubsection{Task 1: Sphero}
The first task was the control of a Sphero---a spherical robot that can roll across the ground at various speeds. In this task, participants had to control the robot around a 5x5 foot course (exemplified in Figure \ref{fig:sphero}). After travelling from spot 1 $\rightarrow$ 2 $\rightarrow$ 3 $\rightarrow$ 4 $\rightarrow$ 1, the task was considered complete. Note that this course had to be modified for two participants: one who suffered from muscle fatigue due to an underlying injury unrelated to the experiment and one who had challenges eliciting a repeatable hand-open contraction. The mapping from gestures to the robot control were Hand Open (Move Away), Hand Close (Move Toward), Wrist Flexion (Move Left), Wrist Extension (Move Right), and No Movement (Do Nothing). The flexion and extension commands were reversed for left-handed users to improve intuitiveness. When prompted, the participants walked over to the course, snapped to turn the robot on, navigated it through the obstacle course, snapped to turn the robot off, and then returned to their respective activities. The only activity the user continued to perform during control was walking on the treadmill.

\subsubsection{Task 2: Alarm Deactivation}
The second task was simpler and involved dismissing an alarm.
Participants snapped to activate the system, extended their wrist to silence the alarm, and then snapped to deactivate the system. Although easier and only requiring a few seconds, this task aimed to simulate a more ubiquitous use case of system control where participants were not required to completely switch contexts. This meant users dismissed the alarm while continuing to perform the activity they were doing (such as while driving) or took a momentary pause (such as when writing).

\subsection{Data Collection, Interview, and Survey}
A set of seven other potential wake gestures was collected at the end of the data collection to evaluate the generalizability of our approach. These gestures were hand open + wrist pronation (pressing a `button'), finger gun (with recoil), hand open (pulse), double middle finger tap, wrist roll (pronation then supination), wrist extension (flick right), and wrist flexion (flick left). Ten repetitions of each gesture were acquired. Finally, participants completed a Likert-style survey, where their experience with the system was assessed alongside supplementary informal interviews. In these unstructured interviews, participants were asked how the system could be improved and their preferences on the tradeoff between true positives and false negatives.

\section{Results}

\subsection{Non Wake Gesture Approach}
\label{sec:non_ww}
\begin{figure*}[h!]
    \centering
    \includegraphics[width=0.8\linewidth]{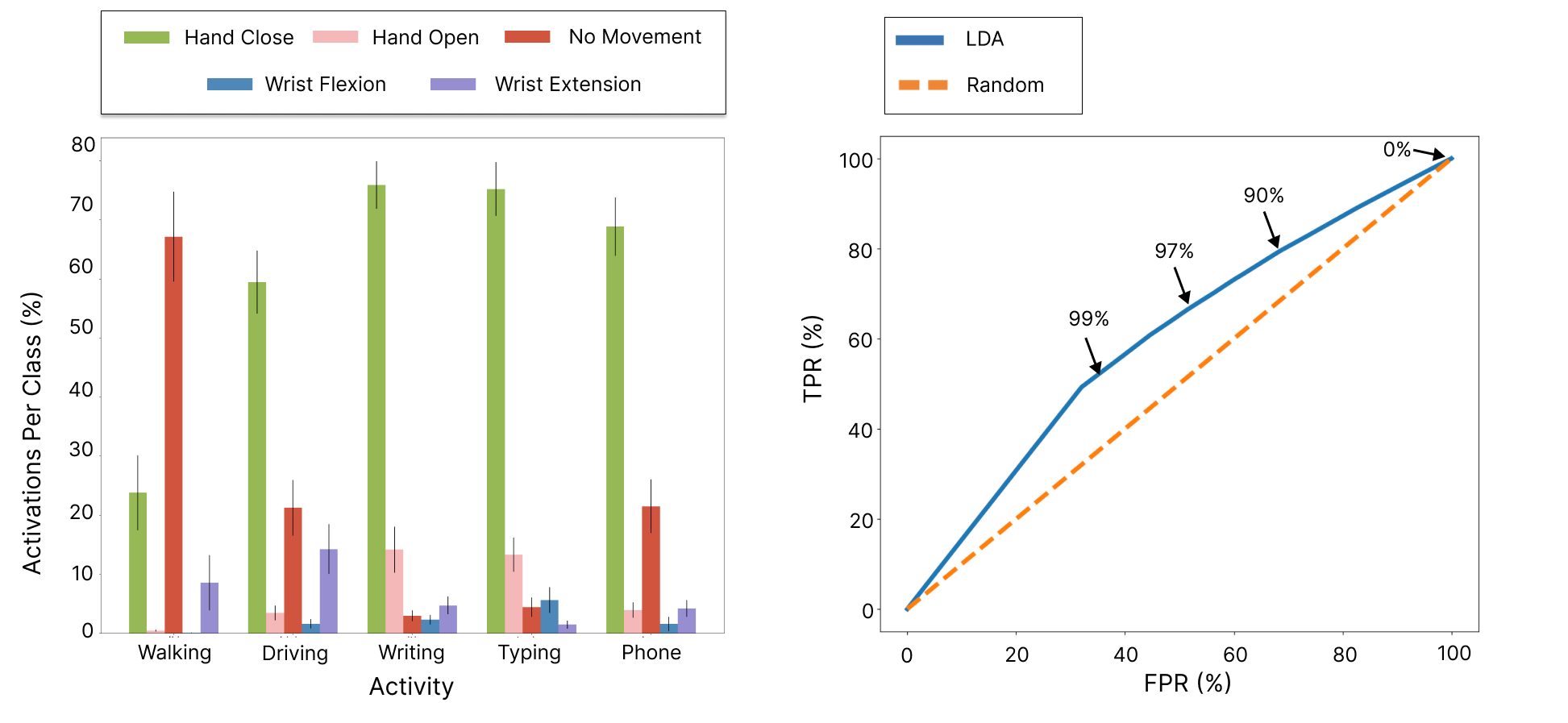}
    \caption{Left: Distribution of class activations averaged across participants during ADLs. Right: The average ROC curve across participants when using confidence-based rejection across a range of threshold values. The TPR is the classifier's performance on the steady-state training data, and the FPR is its ability to reject false activations during ADL tasks recorded during the experiment.}
    \label{fig:rej}
\end{figure*}

The distribution of unintended class activations that would have occurred during the ADLs if no wake gesture was used is shown in Figure \ref{fig:rej} - Left. This was computed post hoc by running the dedicated continuous classifier (outlined in Section \ref{sec:classifier}) on the ADL data recorded when not actively controlling the Sphero during Task 1.  
These results suggest that when performing the ADLs, participants were more often than not eliciting one of the dedicated control gestures. This suggests that a dedicated control system (relying only on the no movement class for inadvertent activations, as is typical in prosthetics) would result in spurious unintended activations if not governed by a wake gesture. This was consistent across most ADLs, with the exception of walking (which required no use of the hands). In theory, this would have meant that the Sphero robot would have been sporadically moving all around the room. 

While the prosthetics community has acknowledged that myoelectric control is an open-set problem through concepts like rejection (i.e., rejecting classifier outputs that fall below a predefined confidence threshold), our results show that such techniques do not work for ignoring out-of-set inputs during ADLs, as highlighted in Figure \ref{fig:rej} - Right. Here, the dedicated continuous classifier's receiver-operating characteristic (ROC) curve is shown if a confidence-based rejection scheme had been used.
This plot indicates that such approaches are only slightly better than randomly guessing if a given input should be rejected.  
These results reinforce the need for specialized strategies to deal with inadvertent muscle-based inputs when leveraging myoelectric control.

\subsection{Wake Gesture System Performance}
\label{sec:proposed}
The number of true positives, false positives, and false negatives were recorded throughout the experiment to evaluate the online performance of the wake gesture system. False positives were counted whenever the system state got toggled, but the user did not snap. In contrast, false negatives were counted whenever an individual elicited a snap gesture, and the system did not get toggled. The number of true negatives (i.e., correctly rejected activity) was determined post hoc by evaluating the total number of decisions made by the wake gesture system on the recorded data from each task. 

\begin{figure}[h!]
    \centering
    \includegraphics[width=\linewidth]{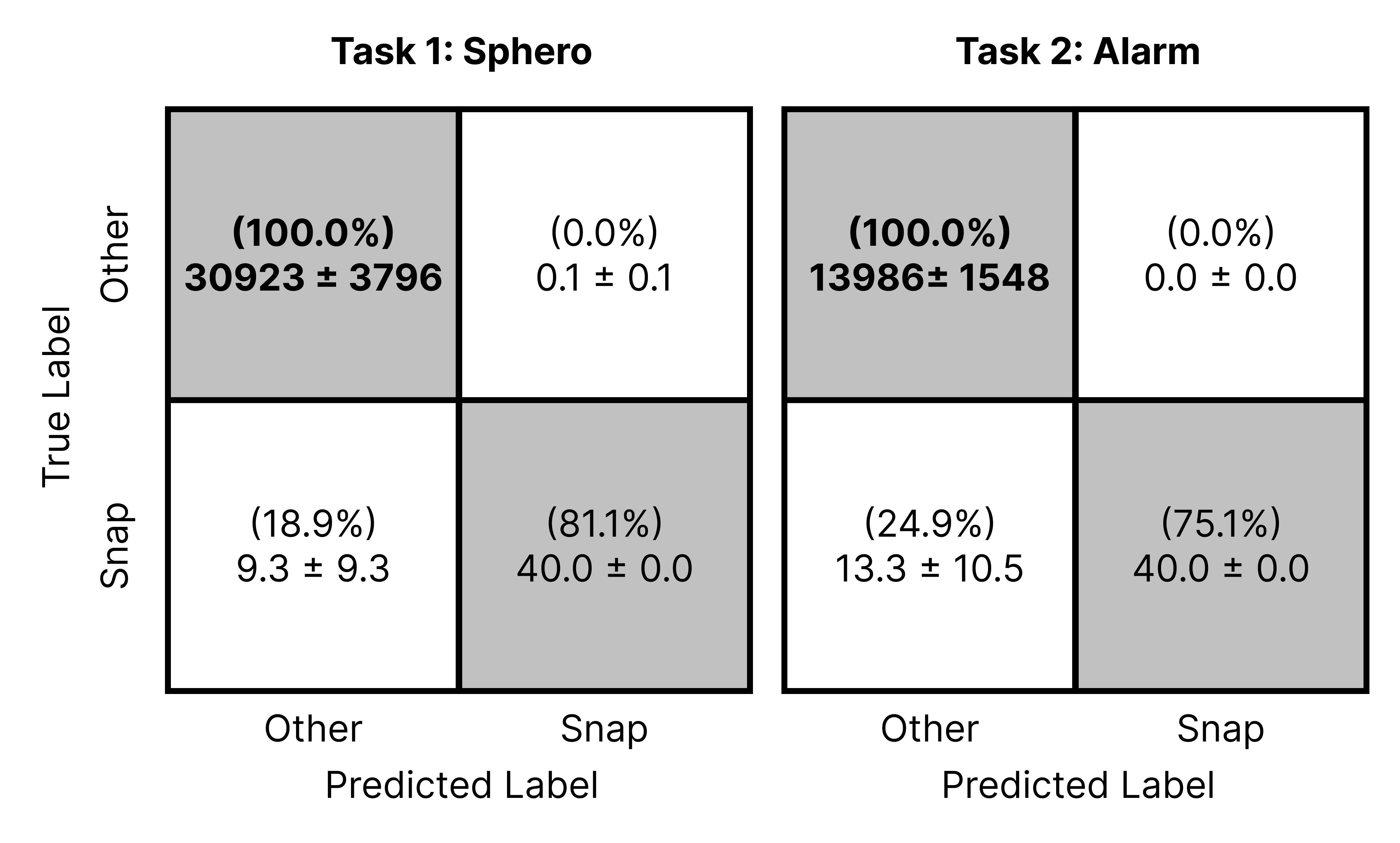}
    \caption{A confusion matrix for each task, showing the predicted and true labels for each decision during the online evaluation of the wake gesture system.}
    \label{fig:conf}
\end{figure}

The confusion matrices presented in Figure \ref{fig:conf} denote the \textit{false negatives} (times that the participant snapped but the system ignored it, bottom left), \textit{false positives} (times that the system was inadvertently activated, top right), \textit{true positives} (correctly identified snaps, bottom right), and \textit{true negatives} (correctly ignored other inputs, top left). The number of true positives was consistent by design across participants and tasks (4 per activity x 5 activities x 2 toggles (one for on and one for off) = 40). The near-zero false positive rate indicates the goal of effectively eliminating inadvertent activations while maintaining an reasonable true positive rate of 75.1\% and 81.1\% for the two respective tasks was achieved. A non-parametric Wilcoxon signed-rank test was run to check for statistical significance between the number of false negatives between the two tasks. The test results revealed a significant difference (z=2.56, p=0.011), indicating that participants had slightly more difficulty replicating the wake gesture template during the alarm task. 

The results of two subjects were withheld from the confusion matrices and their respective analyses as they were deemed outliers. 
The first outlier (z-score $>$ 3) had 141 false negatives, indicating this participant far exceeded the 99.6 percentile of the distribution observed across the other subjects.
Further, there was justification for their removal as they reported having excessive muscle fatigue toward the end of the experiment. This outcome was thought to have been induced by the consistent activation and deactivation of the system, which would be atypical in realistic ubiquitous settings.
The second outlier (z-score $>$ 3) had seven false positives.
This outlier was likely because the threshold for this participant was inadvertently set too high during the initialization of the wake gesture algorithm, failing to properly favour false negatives over false positives. 
It is deduced that this would not have occurred if an adaptive approach that enabled continuous tuning throughout the experiment had been implemented.

\subsection{Evaluating Other Wake Gestures}
\label{sec:o_gestures}
\begin{figure}[h!]
    \centering
    \includegraphics[width=0.85\linewidth]{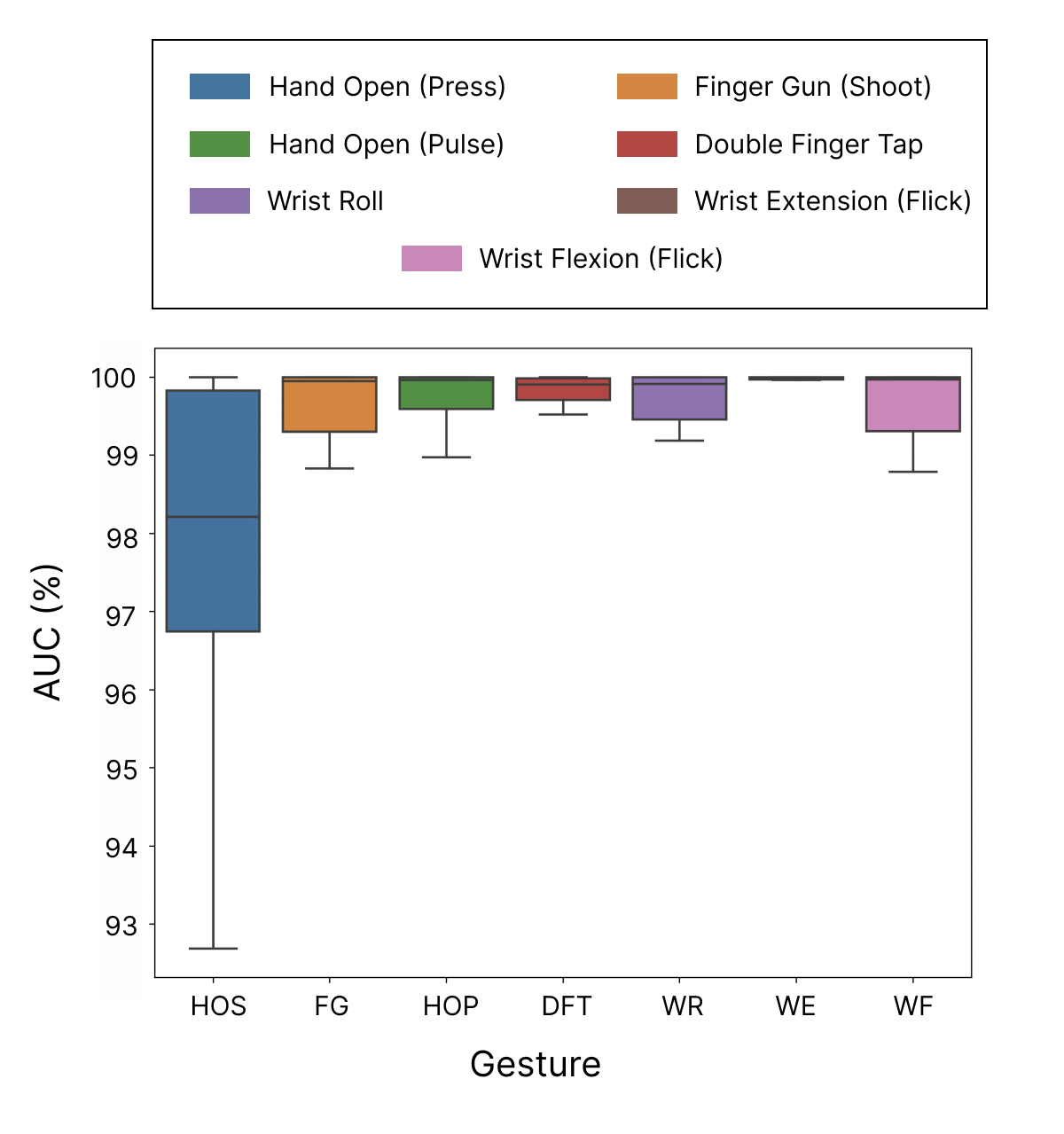}
    \caption{Left: The AUC percentage when using each wake gesture and running the system post hoc on the ADL data. Note that for visualization purposes the outliers have been removed. Right: The average ROC curve for each gesture shows the TPR (its ability to recognize its templates) and the FPR (its ability to ignore the ADL data recorded during the session).}
    \label{fig:gestures}
\end{figure}

The post hoc performance of each of the seven gestures acquired at the end of the study was evaluated by training an offline wake gesture recognition system (as presented in Section \ref{sec:ww_train_sec}) and running it over all of the ADL data acquired during both tasks. Figure \ref{fig:gestures} shows the area under the ROC curve (AUC) percentage across participants. The AUC is the integral of the ROC curve for a single participant, resulting in a value between zero and one, indicating the model's ability to accept deliberate activations and reject inadvertent input. A perfect wake gesture system would have an AUC percentage of 100\%, indicating that it can ignore all ADL data while accepting all the training gesture templates. A non-parametric Friedman test with a Finner posthoc comparison was used to check for significant differences between the AUCs of the different gestures, with the only significant difference being found between the button press and wrist extension gestures (p $<$ 0.005). This highlights the generalizability of this approach to a broader subset of gesture inputs beyond the finger snap used in this study. Note that the finger snap was not included in this analysis because establishing its AUC post hoc isn't feasible due to the presence of deliberate snap wake gestures during the ADL tasks.

\subsection{User Survey and Interview}
\label{sec:sub}
\begin{figure}[h!]
    \centering
    \includegraphics[width=\linewidth]{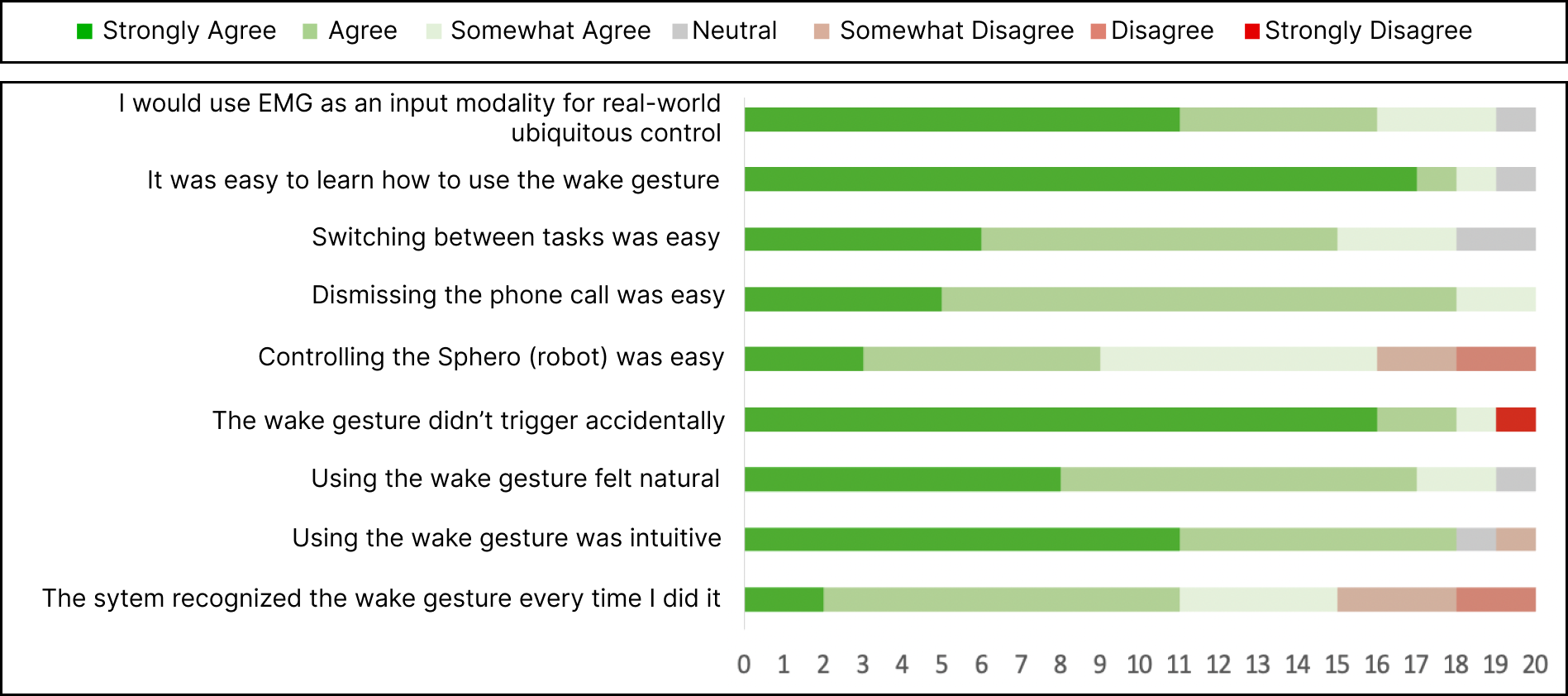}
    \caption{The results from the subjective Likert-style questionnaire evaluating user perception of the on-demand myoelectric control system.}
    \label{fig:quest}
\end{figure}

Finally, to capture participants' subjective feelings on using EMG-based wake gestures, they completed a Likert-style questionnaire, presented in Figure \ref{fig:quest}. In general, the perception of the system was positive, despite the intentional over-rejection of some deliberate activations. The most critical participant (the `strongly disagree' for the accidental triggering of the wake gesture) was the outlier participant mentioned above, who still only had seven false positives in the entire experiment, further highlighting their detriment to user experience. Through informal interviews with participants, there was a consensus that a wake gesture system should over-reject potential activities to ensure no false positives, as dealing with these errors requires more effort from the user. Moreover, it was generally agreed that false negatives were not overly cumbersome, with one participant even noting that having to redo the wake gesture once was not a big deal since snapping is ``so quick and easy''. Another participant described that over-rejecting is like the ``casing on the big red button'' (i.e., it was an added protection against undesirable consequences) emphasizing that they preferred the extra layer of protection against the possibility of inadvertent activations. However, it should be noted that many participants commented that the more false negatives that occurred in succession, the more frustrated they became. Another participant questioned the social acceptability of the snap gesture due to the noise it makes, highlighting the use of other, more discrete wake gestures. In general, however, the consensus was that the snap wake gesture was intuitive and natural, with one participant suggesting that it was an easier way to activate a system than voice-activated smart home systems.

\section{Discussion}
This work proposes, evaluates, and sets a baseline for a novel myoelectric control paradigm---on-demand myoelectric control---for extending the capabilities of muscle-based inputs to ubiquitous environments. The resilience of the proposed wake gesture approach to inadvertent activations was demonstrated during everyday activities, including writing, walking, typing, driving, and phone interaction. A traditional dedicated classification approach would not have worked during these tasks due to the constant unintentional recognition of muscular inputs. 
The proposed template-based approach for recognizing a discrete finger snap to toggle the dedicated control system effectively reduced false activations. In general, it could ignore \textit{all inadvertent activations} during the set of ADLs while maintaining adequate recognition rates of the wake gesture to enable reliable, ubiquitous input. 
While this work provides a solid foundation for the future of on-demand general-purpose myoelectric control, more work is needed to advance myoelectric control to a widely adopted ubiquitous input modality. With this in mind, this section critiques this work and highlights future research directions to better characterize its contributions.

\subsection{The Tension Between False Negatives and False Positives}
For any binary classification problem, the ability of a system to ignore inadvertent activations will always be at odds with its ability to recognize true positives. Improving (or decreasing) the false negative rate will increase the number of false positives and vice versa, unless the underlying separability and repeatability are improved (e.g. better features, more sensors). While the wake gesture system proposed in this study achieved a median false positive rate of 0.0\%, this came at the cost of an increased false negative rate (18.9\% and 24.9\% in the Sphero and alarm tasks, respectively). This meant that participants had to evoke the wake gesture approximately one out of five times more than once. Although seemingly undesirable, post-study interviews with participants confirmed that these false negatives were much preferred to inadvertent activations, corroborating previous work evaluating the tradeoff between the two \cite{fn_v_fp}. Interestingly, this was noted even though 18/20 participants were not impacted by false positives during the experiment, as the control schemes were deliberately tuned this way. 
A significant pain point discussed by many of the participants, however, was when they were required to repeat the wake gesture more than a few times. This highlights the need for user-adjustable thresholds or adaptive algorithms that actively detect successive intentional activations (and potentially temporarily increase the threshold), reducing the number of consecutive false negatives. Additionally, future work could improve the system online by incorporating new templates that lead to a successful interaction session, continuously adapting the system. Generally, however, now that a baseline has been set, future work should focus on increasing the true positive rate while maintaining a false positive rate of 0\%.

\subsection{Other Applications for Wake Gestures}
One seemingly promising use of wake gestures could be during prosthesis control, where there is still a high abandonment rate of myoelectric prostheses \cite{abandonment}. Such a wake gesture approach would enable amputees to selectively toggle their prostheses into an active state when desired and an inactive state when performing activities in which their prostheses are not needed. Beyond simply an on/off switch, amputees could lock their prosthesis in specific grasps without worrying about spurious activations (such as when holding a warm cup of coffee). Future work should evaluate the efficacy of wake gestures in these prosthesis control settings.

Beyond myoelectric control, this work is generalizable to any ubiquitous input modality in which false activations could occur. Novel input modalities, including myoelectric control, are often evaluated in highly constrained settings. However, any interaction techniques that rely on sensing modalities that do not have a general default inactive state (i.e., that have to worry about inadvertent activations) will struggle to find adoption within real-world use cases. For example, although researchers can achieve high recognition accuracy on IMU-based gesture recognition tasks like air-drawing \cite{imu_input}, how can these systems be enabled in real-world scenarios when using our hands for other tasks? As highlighted in this work, optimizing offline recognition and evaluating these modalities in constrained settings is inadequate if the goal is to progress these technologies toward widespread consumer adoption.

\subsection{Algorithmic Improvements and Advancements}
This work evaluated a simple time series distance measurement (DTW) to measure the distance between target templates and potential wake gestures. The issue, however, is that this approach scales linearly with the number of templates and does not lend well to cross-user models or reuse across sessions. Future work should explore other algorithms (e.g., deep learning) or more robust feature sets to improve the resilience of wake gesture systems to such factors. However, one interesting thing to note is the importance that DTW played in considering the temporal profile of the gesture rather than traditional machine learning approaches that assume stationarity. For example, when the temporal structure of wrist flicks was leveraged (extension and flexion), they could be actively discerned from the ADL data (i.e., the AUC was high). This was not the case when recognizing the static contractions of flexion and extension, where the LDA classifier could not differentiate these non-dynamic inputs from the ADLs. These results highlight a promising path forward for using dynamic gestures within HCI applications \cite{eddy_hci}.

Other modalities, such as IMUs, will often be available when considering wearable devices, lending to potentially more robust opportunities for wake gesture detection. Combining modalities for wake gesture recognition is an interesting area for future work. Another outcome was the significant difference in TPR between the Sphero and Alarm tasks. It is conceivable that without switching contexts, users were more distracted when eliciting the wake gesture, thus leading to less repeatable gestures. Future work should consider how to train algorithms to become resilient to such confounding factors through real-time adaptation or more representative data collection. Finally, although the proposed approach could not be compared against the original Myo Armband control scheme due to the removal of all documentation and software from the web, it has been nearly ten years since its discontinuation. In this time, massive improvements have been made to EMG hardware and in machine learning. In turn, the research community is more equipped than ever to solve the challenges faced by the Myo Armband, particularly in unlocking EMG as a widely applicable ubiquitous input. 

\section{Conclusion}
This work takes a meaningful step toward the real-world viability of myoelectric control as a widely applicable ubiquitous input. In particular, the results suggest that wake gestures are a viable solution for enabling on-demand myoelectric control, effectively eliminating inadvertent activations during everyday ubiquitous tasks. In an online evaluation, near-perfect performance in rejecting unintended device inputs while maintaining adequate robustness for two myoelectric control tasks is achieved (dismissing an alarm and controlling a robot). Moving forward, researchers should consider using wake gestures when designing interactive systems, particularly when using sensed inputs that may be inadvertently activated during unrelated activities of daily living.

\bibliographystyle{ieeetr}
\bibliography{bib}

\begin{thebibliography}{10}

\bibitem{21_cent_comp}
M.~Weiser, ``The computer for the 21st century,'' {\em SIGMOBILE Mob. Comput. Commun. Rev.}, vol.~3, p.~3–11, jul 1999.

\bibitem{emg_hms}
W.~Guo, X.~Sheng, H.~Liu, and X.~Zhu, ``Toward an enhanced human–machine interface for upper-limb prosthesis control with combined emg and nirs signals,'' {\em IEEE Transactions on Human-Machine Systems}, vol.~47, no.~4, pp.~564--575, 2017.

\bibitem{eddy_hci}
E.~Eddy, E.~J. Scheme, and S.~Bateman, ``A framework and call to action for the future development of emg-based input in hci,'' in {\em Proceedings of the 2023 CHI Conference on Human Factors in Computing Systems}, CHI '23, (New York, NY, USA), Association for Computing Machinery, 2023.

\bibitem{scheme_electromyogram_2011}
E.~Scheme and K.~Englehart, ``Electromyogram pattern recognition for control of powered upper-limb prostheses: state of the art and challenges for clinical use,'' {\em Journal of Rehabilitation Research and Development}, vol.~48, no.~6, pp.~643--659, 2011.

\bibitem{oldest_prosthetic1}
J.~Baits, R.~Todd, and J.~Nightingale, ``Paper 10: The feasibility of an adaptive control scheme for artificial prehension,'' in {\em Proceedings of the Institution of Mechanical Engineers, Conference Proceedings}, vol.~183, pp.~54--59, SAGE Publications Sage UK: London, England, 1968.

\bibitem{saponas_demonstrating_2008}
T.~S. Saponas, D.~S. Tan, D.~Morris, and R.~Balakrishnan, ``Demonstrating the feasibility of using forearm electromyography for muscle-computer interfaces,'' in {\em Proceedings of the {SIGCHI} {Conference} on {Human} {Factors} in {Computing} {Systems}}, {CHI} '08, (New York, NY, USA), pp.~515--524, Association for Computing Machinery, Apr. 2008.

\bibitem{imprecise_but_fun}
J.~Karolus, S.~Thanheiser, D.~Peterson, N.~Viot, T.~Kosch, A.~Schmidt, and P.~W. Wozniak, ``Imprecise but fun: Playful interaction using electromyography,'' {\em Proc. ACM Hum.-Comput. Interact.}, vol.~6, sep 2022.

\bibitem{continuous_control}
K.~Englehart and B.~Hudgins, ``A robust, real-time control scheme for multifunction myoelectric control,'' {\em IEEE Transactions on Biomedical Engineering}, vol.~50, no.~7, pp.~848--854, 2003.

\bibitem{libemg}
E.~Eddy, E.~Campbell, A.~Phinyomark, S.~Bateman, and E.~Scheme, ``Libemg: An open source library to facilitate the exploration of myoelectric control,'' {\em IEEE Access}, vol.~11, pp.~87380--87397, 2023.

\bibitem{drone}
Y.~Doshi and D.~Nath, ``Designing a drone controller using electromyography signals,'' in {\em 2021 International Conference on Communication information and Computing Technology (ICCICT)}, pp.~1--6, 2021.

\bibitem{smart_garments}
S.~Benatti, E.~Farella, and L.~Benini, ``Towards emg control interface for smart garments,'' in {\em Proceedings of the 2014 ACM International Symposium on Wearable Computers: Adjunct Program}, ISWC '14 Adjunct, (New York, NY, USA), p.~163–170, Association for Computing Machinery, 2014.

\bibitem{dtw}
H.~Sakoe and S.~Chiba, ``Dynamic programming algorithm optimization for spoken word recognition,'' {\em IEEE Transactions on Acoustics, Speech, and Signal Processing}, vol.~26, no.~1, pp.~43--49, 1978.

\bibitem{direct_control}
M.~M. White, W.~Zhang, A.~T. Winslow, M.~Zahabi, F.~Zhang, H.~Huang, and D.~B. Kaber, ``Usability comparison of conventional direct control versus pattern recognition control of transradial prostheses,'' {\em IEEE Transactions on Human-Machine Systems}, vol.~47, no.~6, pp.~1146--1157, 2017.

\bibitem{abandonment}
K.~Østlie, I.~M. Lesjø, R.~J. Franklin, B.~Garfelt, O.~H. Skjeldal, and P.~Magnus, ``Prosthesis rejection in acquired major upper-limb amputees: a population-based survey,'' {\em Disability and Rehabilitation: Assistive Technology}, vol.~7, no.~4, pp.~294--303, 2012.
\newblock PMID: 22112174.

\bibitem{out-of-set-emg}
L.~Wu, X.~Zhang, X.~Zhang, X.~Chen, and X.~Chen, ``Metric learning for novel motion rejection in high-density myoelectric pattern recognition,'' {\em Knowledge-Based Systems}, vol.~227, p.~107165, 2021.

\bibitem{rejection}
E.~J. Scheme, B.~S. Hudgins, and K.~B. Englehart, ``Confidence-based rejection for improved pattern recognition myoelectric control,'' {\em IEEE Transactions on Biomedical Engineering}, vol.~60, no.~6, pp.~1563--1570, 2013.

\bibitem{rejection2}
T.~Bao, S.~A.~R. Zaidi, S.~Q. Xie, P.~Yang, and Z.-Q. Zhang, ``Cnn confidence estimation for rejection-based hand gesture classification in myoelectric control,'' {\em IEEE Transactions on Human-Machine Systems}, vol.~52, no.~1, pp.~99--109, 2022.

\bibitem{pen_agumentation}
F.~Matulic, B.~Vogel, N.~Kimura, and D.~Vogel, ``Eliciting pen-holding postures for general input with suitability for emg armband detection,'' in {\em Proceedings of the 2019 ACM International Conference on Interactive Surfaces and Spaces}, ISS '19, (New York, NY, USA), p.~89–100, Association for Computing Machinery, 2019.

\bibitem{plug-and-play}
J.-i. Furukawa, S.~Chiyohara, T.~Teramae, A.~Takai, and J.~Morimoto, ``A collaborative filtering approach toward plug-and-play myoelectric robot control,'' {\em IEEE Transactions on Human-Machine Systems}, vol.~51, no.~5, pp.~514--523, 2021.

\bibitem{campbell2020feasibility}
E.~Campbell, J.~A. Cameron, and E.~Scheme, ``Feasibility of data-driven emg signal generation using a deep generative model,'' in {\em 2020 42nd Annual International Conference of the IEEE Engineering in Medicine \& Biology Society (EMBC)}, pp.~3755--3758, IEEE, 2020.

\bibitem{campbell2021deep}
E.~Campbell, A.~Phinyomark, and E.~Scheme, ``Deep cross-user models reduce the training burden in myoelectric control,'' {\em Frontiers in Neuroscience}, vol.~15, p.~657958, 2021.

\bibitem{midas_touch}
R.~J.~K. Jacob, ``The use of eye movements in human-computer interaction techniques: What you look at is what you get,'' {\em ACM Trans. Inf. Syst.}, vol.~9, p.~152–169, apr 1991.

\bibitem{jason_adl}
J.~Chang, A.~Phinyomark, S.~Bateman, and E.~Scheme, ``Wearable emg-based gesture recognition systems during activities of daily living: An exploratory study,'' in {\em 2020 42nd Annual International Conference of the IEEE Engineering in Medicine \& Biology Society (EMBC)}, pp.~3448--3451, 2020.

\bibitem{speech_wakewords}
H.~Jung and H.~Kim, ``Finding contextual meaning of the wake word,'' in {\em Proceedings of the 1st International Conference on Conversational User Interfaces}, CUI '19, (New York, NY, USA), Association for Computing Machinery, 2019.

\bibitem{rf-sensor-ww}
E.~Kurtoğlu, A.~C. Gurbuz, E.~A. Malaia, D.~Griffin, C.~Crawford, and S.~Z. Gurbuz, ``Asl trigger recognition in mixed activity/signing sequences for rf sensor-based user interfaces,'' {\em IEEE Transactions on Human-Machine Systems}, vol.~52, no.~4, pp.~699--712, 2022.

\bibitem{ww_pradeep}
P.~Kumar, A.~Phinyomark, and E.~Scheme, ``Verification-based design of a robust emg wake word,'' in {\em 2021 43rd Annual International Conference of the IEEE Engineering in Medicine \& Biology Society (EMBC)}, pp.~638--642, 2021.

\bibitem{tavakoli_1}
M.~Tavakoli, C.~Benussi, and J.~L. Lourenco, ``Single channel surface emg control of advanced prosthetic hands: A simple, low cost and efficient approach,'' {\em Expert Systems with Applications}, vol.~79, pp.~322--332, 2017.

\bibitem{tavakoli_2}
M.~Tavakoli, C.~Benussi, P.~{Alhais Lopes}, L.~B. Osorio, and A.~T. {de Almeida}, ``Robust hand gesture recognition with a double channel surface emg wearable armband and svm classifier,'' {\em Biomedical Signal Processing and Control}, vol.~46, pp.~121--130, 2018.

\bibitem{myo_double_tap}
M.~Cognolato, M.~Atzori, D.~Faccio, C.~Tiengo, F.~Bassetto, R.~Gassert, and H.~Muller, ``Hand gesture classification in transradial amputees using the myo armband classifier,'' in {\em 2018 7th IEEE International Conference on Biomedical Robotics and Biomechatronics (Biorob)}, pp.~156--161, 2018.

\bibitem{locked_out}
C.~H. Fong, M.~Billinghurst, Z.~S. See, and H.~Esmaeili, ``Peppergram with interactive control,'' in {\em 2016 22nd International Conference on Virtual System \& Multimedia (VSMM)}, pp.~1--5, 2016.

\bibitem{wake_gesture_problematic}
F.~Kerber, P.~Lessel, and A.~Kr\"{u}ger, ``Same-side hand interactions with arm-placed devices using emg,'' in {\em Proceedings of the 33rd Annual ACM Conference Extended Abstracts on Human Factors in Computing Systems}, CHI EA '15, (New York, NY, USA), p.~1367–1372, Association for Computing Machinery, 2015.

\bibitem{aaron_momo}
A.~Tabor, S.~Bateman, E.~Scheme, D.~R. Flatla, and K.~Gerling, ``Designing game-based myoelectric prosthesis training,'' in {\em Proceedings of the 2017 CHI Conference on Human Factors in Computing Systems}, CHI '17, (New York, NY, USA), p.~1352–1363, Association for Computing Machinery, 2017.

\bibitem{fn_v_fp}
B.~Lafreniere, T.~R.~Jonker, S.~Santosa, M.~Parent, M.~Glueck, T.~Grossman, H.~Benko, and D.~Wigdor, ``False positives vs. false negatives: The effects of recovery time and cognitive costs on input error preference,'' in {\em The 34th Annual ACM Symposium on User Interface Software and Technology}, UIST '21, (New York, NY, USA), p.~54–68, Association for Computing Machinery, 2021.

\bibitem{snap_motivation}
P.~Pomykalski, M.~P. Wo\'{z}niak, P.~W. Wo\'{z}niak, K.~Grudzie\'{n}, S.~Zhao, and A.~Romanowski, ``Considering wake gestures for smart assistant use,'' in {\em Extended Abstracts of the 2020 CHI Conference on Human Factors in Computing Systems}, CHI EA '20, (New York, NY, USA), p.~1–8, Association for Computing Machinery, 2020.

\bibitem{prosthesis_guided_training}
C.~L. Chicoine, A.~M. Simon, and L.~J. Hargrove, ``Prosthesis-guided training of pattern recognition-controlled myoelectric prosthesis,'' in {\em 2012 Annual International Conference of the IEEE Engineering in Medicine and Biology Society}, pp.~1876--1879, 2012.

\bibitem{campbell2019linear}
E.~Campbell, A.~Phinyomark, and E.~Scheme, ``Linear discriminant analysis with bayesian risk parameters for myoelectric control,'' in {\em 2019 IEEE Global Conference on Signal and Information Processing (GlobalSIP)}, pp.~1--5, IEEE, 2019.

\bibitem{hudgins}
B.~Hudgins, P.~Parker, and R.~Scott, ``A new strategy for multifunction myoelectric control,'' {\em IEEE Transactions on Biomedical Engineering}, vol.~40, no.~1, pp.~82--94, 1993.

\bibitem{motion_normalized}
E.~Scheme, B.~Lock, L.~Hargrove, W.~Hill, U.~Kuruganti, and K.~Englehart, ``Motion normalized proportional control for improved pattern recognition-based myoelectric control,'' {\em IEEE Transactions on Neural Systems and Rehabilitation Engineering}, vol.~22, no.~1, pp.~149--157, 2014.

\bibitem{imu_input}
J.~Younas, H.~Margarito, S.~Bian, and P.~Lukowicz, ``Finger air writing - movement reconstruction with low-cost imu sensor,'' in {\em MobiQuitous 2020 - 17th EAI International Conference on Mobile and Ubiquitous Systems: Computing, Networking and Services}, MobiQuitous '20, (New York, NY, USA), p.~69–75, Association for Computing Machinery, 2021.

\end{thebibliography}

\end{document}